\documentclass[usenatbib]{mnras}
\usepackage{graphicx}
\usepackage{color}
\usepackage{setspace}
\usepackage{dcolumn}
\newcolumntype{d}[1]{D{.}{.}{#1}}

\makeatletter
\newlength{\abovecaptionskip}%
\makeatother
\usepackage{threeparttable}

\usepackage{multicol}
\usepackage{pdflscape}
\usepackage{afterpage}

\def\apj{\rm ApJ}
\def\apjl{\rm ApJL}
\def\apjs{\rm ApJS}

\def\mnras{\rm MNRAS}
\def\nat{\rm Nature}

\def\aap{\rm AAP}
\def\araa{\rm ARA\&A}

\def\gax{\mathrel{\raise.3ex\hbox{$>$}\mkern-14mu\lower0.6ex\hbox{$\sim$}}}
\def\lax{\mathrel{\raise.3ex\hbox{$<$}\mkern-14mu\lower0.6ex\hbox{$\sim$}}}
\def\gtorder{\mathrel{\raise.3ex\hbox{$>$}\mkern-14mu
             \lower0.6ex\hbox{$\sim$}}}
\def\ltorder{\mathrel{\raise.3ex\hbox{$<$}\mkern-14mu
             \lower0.6ex\hbox{$\sim$}}}

\begin{document}

\title[Light Curve Constraints in Quasar Microlensing]{Quasar Microlensing Models with Constraints on the Quasar Light Curves}

\author[Tie \& Kochanek]{ 
    S.~S. Tie$^1$,
    C.~S. Kochanek$^{1,2}$
    \\
  $^{1}$ Department of Astronomy, The Ohio State University, 140 West 18th Avenue, Columbus OH 43210 \\
  $^{2}$ Center for Cosmology and AstroParticle Physics, The Ohio State University,
    191 W. Woodruff Avenue, Columbus OH 43210 \\
   }

\maketitle

\begin{abstract}
Quasar microlensing analyses implicitly generate a model of the variability of the source quasar. The implied source variability may be unrealistic yet its likelihood is generally not evaluated. We used the damped random walk (DRW) model for quasar variability to evaluate the likelihood of the source variability and applied the revised algorithm to a microlensing analysis of the lensed quasar RX J1131$-$1231. We compared the estimates of the source quasar disk and average lens galaxy stellar mass with and without applying the DRW likelihoods for the source variability model and found no significant effect on the estimated physical parameters. The most likely explanation is that unreliastic source light curve models are generally associated with poor microlensing fits that already make a negligible contribution to the probability distributions of the derived parameters.
\end{abstract}


\section{Introduction}
\label{sec:intro}
Quasar accretion disks cannot be resolved by either ground- or space-based telescopes. This has led to the development of methods that estimate the size of the emitting region without resolving it, such as reverberation mapping of the quasar continuum emission (e.g., \citealp{Collier1998,Sergeev2005,Shappee2014,Fausnaugh2016}), modeling of the structure of the broad Fe K$\alpha$ line (e.g., \citealp{Tanaka1995,Iwasawa1996,Fabian2002,Iwasawa2004}), and gravitational microlensing (e.g.. \citealp{Pooley2007,Morgan2008,Poindexter2008,Dai2010,Blackburne2011,Chartas2012,Mosquera2013,MacLeod2015}). In microlensing, stars near the images of gravitationally lensed quasars produce time variable magnifications. The amplitude of the microlensing variability depends on the relative sizes of the quasar accretion disk, $R_{S}$, and the star's Einstein radius, $R_{E}$ (see the review by \citealt{Wambsganss2006}). A larger disk size $R_{S}$ relative to $R_{E}$ smooths out the microlensing effect and results in smaller variability amplitudes, and vice versa. In essence, the amplitude of the microlensing variability constrains the size of the source. 

\cite{Kochanek2004} (hereafter K04) and \cite{Poindexter2010} developed a general Monte-Carlo method for analyzing quasar microlensing light curves to simultaneously estimate the size of the source as well as properties of the lens galaxy, such as its average stellar mass and its dark matter content. The basic procedure is to first produce random microlensing magnification patterns for a range of lens models consistent with the observed properties of the lens, convolve these patterns with a model for the quasar brightness profile, and then generate trial light curves by randomly selecting source trajectories across the patterns. The random trial light curves are then compared to the observed light curves using $\chi^{2}$ statistics. Finally, Bayesian methods are used to combine the results of all the trials into estimates of the model parameters. Each trial light curve implicitly makes an estimate of the intrinsic variability of the source quasar based on the data and the microlensing model, but the original algorithm has no means of evaluating whether this model for the intrinsic variability is statistically likely for a quasar. In other words, the algorithm includes no likelihood for the source light curve. 

Recently, it has been found that the optical variability of quasars can be reasonably well modeled as a damped random walk (DRW) stochastic process \citep{Kelly2009,Kozlowski2010,MacLeod2010}. The DRW process is described by an exponential covariance matrix $S$ of the signal between epochs $t_{i}$ and $t_{j}$,
\begin{equation}
S_{ij} = \sigma^{2}\mbox{exp}(-|t_{i}-t_{j}|/\tau),
\end{equation}
\noindent
 where $\tau$ is the time scale for the light curve to de-correlate and $\sigma$ describes the overall variability amplitude. The parameters are correlated with wavelength and the properties of the quasar \citep{MacLeod2010} and they may physically be driven by thermal instabilities \citep{Kelly2009}. While there may be deviations from the DRW model on short time scales (see, e.g., \citealp{Mushotzky2011,Zu2013,Kasliwal2015,Kozlowski2016}), the model provides a good statistical description of quasar variability on time scales of weeks to decades. 

In the K04 algorithm, the source light curve model ``absorbs'' as much of the difference between the true microlensing variability of the data and that of the trial light curve as it is statistically allowed (see Eqn. 5 in K04). If these differences are large, it may imply intrinsic variability that is atypical of a quasar and so should be poorly fit as a DRW. In this paper, we test this hypothesis by including the DRW model into the initial K04 algorithm  to evaluate the likelihood of the source model for the intrinsic variability of the quasar as part of the parameter estimation process. We then applied the modified algorithm to RX~J1131$-$1231, a $z_s = 0.658$ quasar lensed into four images by a $z_l = 0.295$ galaxy \citep{Sluse2003}. This lensed quasar has been the target of many microlensing studies to probe its optical and X-ray emission regions (e.g., \citealp{r1131_blackburne2006,r1131_chartas2009,Dai2010,r1131_chartas2016,r1131_neronov2016}), as well as studies to measure its time-delays (e.g., \citealt{r1131_morgan2006,Tewes2013}). We describe the data and our methods in \S\ref{sec:data}, the results in \S\ref{sec:results}, and a summary of our conclusions in \S\ref{sec:conclusion}.

\section{Data and Methods}
\label{sec:data}
We used the $R$ band optical light curves for the four lensed images of the quasar from \cite{Tewes2013}. The light curves span nine observing seasons from 2004 to 2012 with 707 total epochs and a median spacing of two days. The median single-epoch errors are 0.005, 0.007, 0.01, and 0.027 mag for the A, B, C, and D images, respectively. We checked whether these error estimates are accurate by fitting a straight line to each successive triplet of data points separated by less than 15 days and calculating the goodness of fit. These fits showed significant deviations from a $\chi^{2}$ per degree of freedom (dof) of unity, so we added additional systematic errors in quadrature to the original errors in order to bring the mean of the triplet $\chi^{2}/\mbox{dof}$ to unity. These systematic error estimates are 0.012, 0.02, 0.037, and 0.038 mag for images A$-$D, respectively. 

\begin{figure}
	\centering
    \includegraphics[width=0.45\textwidth]{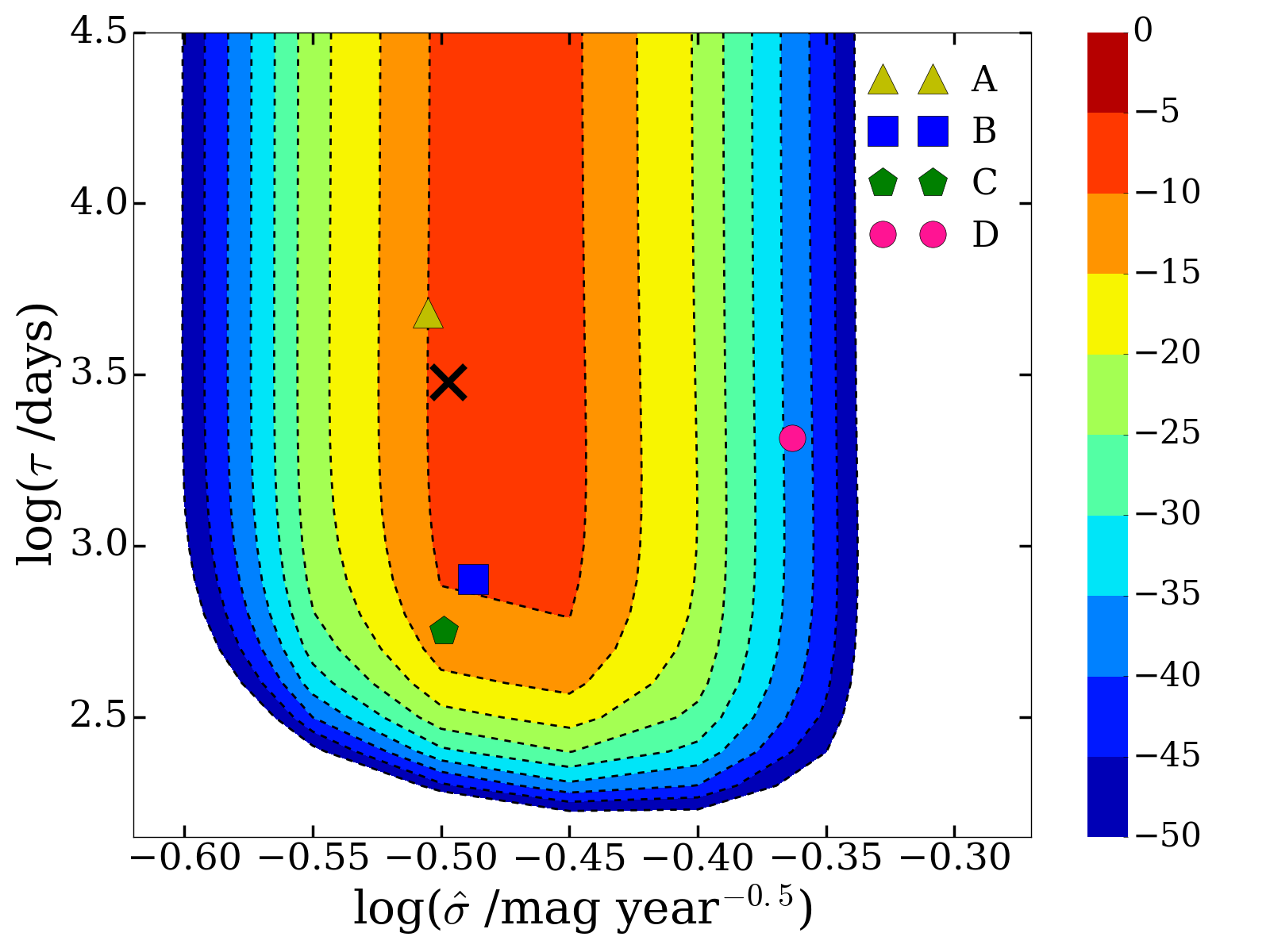}
    \caption{Combined likelihood contour $\prod_{k} P(D_{k}|\hat{\sigma}, \tau)$ for the light curves $k =$ A, B, C, and D shown here in logarithmic scale for the range of trial DRW parameters. The dots are the best-fit DRW parameters for each image and the cross shows the parameters eventually used for the final results ($\hat{\sigma}_{int}$ and $\tau_{int}$ in text). The best-fit parameters for each image are located at the peak and marked but the individual probability distributions are not shown. }
    \label{drwcont}
\end{figure}

We fit the DRW model to each of the individual light curves after adding the systematic errors in quadrature to the original light curves. We used the algorithm from \cite{Kozlowski2010} to obtain the best-fit parameters by maximizing the likelihood of the data given the trial DRW parameters $P(D|\hat{\sigma}, \tau)$. Following \cite{Kozlowski2010}, we used $\hat{\sigma} = (2\sigma^2/\tau)^{1/2}$ as our model parameter instead of $\sigma$ as it is well-constrained even when $\sigma$ and $\tau$ are degenerate. We used a uniform 2D grid of $\hat{\sigma}$ and $\tau$ values ranging from log$_{10}$($\hat{\sigma}$) = $-3.0$ to 1.0 in 0.05 dex increments and log$_{10}$($\tau$) = $-1.0$ to 5.0 in 0.1 dex increments. Our best-fit DRW parameters for each image are given in Table \ref{tab:drw}. Figure \ref{drwcont} shows the combined likelihood from the DRW fitting process, $\prod_{k} P_{k}(D_{k}|\hat{\sigma}, \tau) = P_{A}(D_{A}|\hat{\sigma}, \tau) \times P_{B}(D_{B}|\hat{\sigma}, \tau) \times P_{C}(D_{C}|\hat{\sigma}, \tau) \times P_{D}(D_{D}|\hat{\sigma}, \tau)$, where the dots show the maximum likelihood parameters for each image. 

\begin{figure*}
	\centering
    \includegraphics[width=0.95\textwidth]{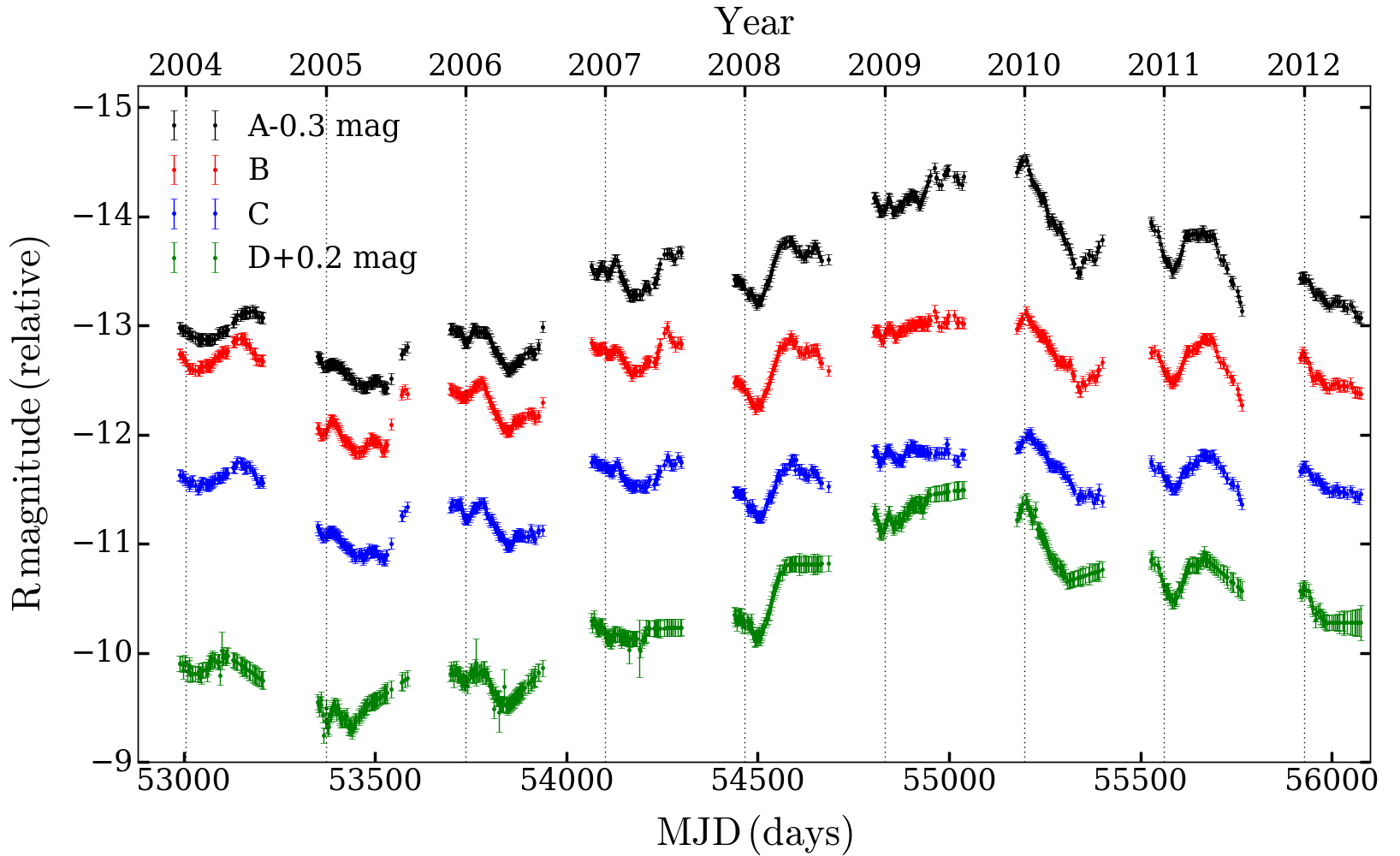}
    \caption{Lensed $R$ band light curves of RX J1131$-$1231 after removing the time delays obtained from \protect\cite{Tewes2013}, interpolating with the DRW model, and averaging points separated by less than 1.5 days. We keep only data where all four shifted light curves overlap.}
    \label{fig:lcfnal}
\end{figure*}

We removed the time delays between the quasar images using the delays obtained by \cite{Tewes2013} and treating image A as our reference image. While the time delays between image A, B, and C are small, the time delay between images A and D is significantly larger, at t$_{A}$ $-$ t$_{D}$ = 90.6 days. We interpolated the light curves in each observing season using the DRW model with $\tau_{int}$ = 3000 days and $\hat{\sigma}_{int}$ = 0.32 mag/$\sqrt{\mbox{year}}$, where $\hat{\sigma}_{int}$ is the average $\hat{\sigma}$ for ABC. The $\hat{\sigma}$ value for D was not included, as it is relatively different from the rest due to the larger variability contribution from microlensing. We fixed $\tau_{int}$ to 3000 days, which is roughly between $\tau_{A}$ and $\tau_{D}$. Changing $\tau$ will have little effect since the joint likelihood spans a large range of $\tau$ as shown in Figure~\ref{drwcont} and estimate of $\tau$ are generally very uncertain (e.g., \citealt{Kozlowski2010,MacLeod2010,Kozlowski2017}). Finally, to speed up our microlensing analyses, we averaged any interpolated points that are separated by less than 1.5 days, weighted by their uncertainties. Figure \ref{fig:lcfnal} shows the interpolated and averaged light curves with the time delays removed. 

\begin{table}
\small
\caption{Best fit DRW parameters} 
\centering           
\begin{tabular}{ccc}
\hline\hline 
Image & $\hat{\sigma}$ (mag/$\sqrt[]{\mbox{year}}$) & $\tau$ (days)\\
%\mbox{Image} & \hat{\sigma} (\mbox{mag}/\sqrt[]{\mbox{year}}) & \tau (\mbox{days})\\
\hline  
A & 0.312 & 4794.9\\[0.5ex]
B & 0.326 & \ \ 799.8\\[0.5ex]
C & 0.317 & \ \ 566.9\\[0.5ex]
D & 0.433 & 2068.6\\[0.5ex]
\hline
\end{tabular}
\label{tab:drw}
\end{table}

We used the methods of K04 for our microlensing analyses and the lens models from \cite{Dai2010}. Specifically, the lens was first fitted using an ellipsoidal $R_{e} = 1\farcs7$ de Vaucouleurs (deV) profile for the stellar mass distribution. We then added an NFW dark matter halo, considering a range of lens models with different stellar mass fractions $f_{*}$ relative to a pure deV model from $f_{*} = 0.1$ to $f_{*} = 1$ in increments of $\Delta f_{*} = 0.1$, where $f_{*} = 1$ is the pure deV model. We used static magnification patterns and did not consider the random motion of the stars in the lens galaxy, which avoided the need to parallelize the DRW likelihood calculations (but see \citealp{Poindexter2010}). We generated eight random realizations of the magnification patterns for each lens model based on the microlensing parameters, convergence $\kappa$, shear $\gamma$, and fraction of stellar convergence $\kappa_{*}/\kappa$, from \cite{Dai2010}. Each magnification pattern has 8192 $\times$ 8192 pixels, an outer scale of 10$\langle R_{e} \rangle$, and an inner pixel scale of 10$\langle R_{e} \rangle$/8192 $\sim$ 0.001$\langle R_{e} \rangle$. Following \cite{Dai2010} and the velocity model from K04, the projection of the cosmic microwave background dipole velocity on the lens plane and the lens velocity dispersion are taken to be 47 km s$^{-1}$ and 350 km s$^{-1}$, respectively. Similarly, we used $235/(1 + z_{l})$ km/s $\approx$ 180 km/s and $235/(1 + z_{s})$ km/s $\approx$ 140 km/s for the rms peculiar velocities of the lens and source galaxies, respectively. 

We modeled the source (the accretion disk) as a circular, face-on, standard thin disk \citep{Shakura1973}. In generating the trial light curves, we used a logarithmic grid of source sizes log$_{10}$($R_{S}$/cm) from 14.50 to 17.50 with a spacing of 0.1 dex. Each simulated light curve at epoch $i$ consists of the source magnitude $S_{i}$ and the total logarithmic magnification $\mu_{tot,i} = \mu + \Delta\mu + \delta\mu_{i}$, comprising the magnification $\mu$ of the ``macro'' model, any offset $\Delta\mu$ from that model, and the micro magnification $\delta\mu_{i}$. Thus, the light curve of image $\alpha$ is modeled as 
\begin{equation}
m^{\alpha}_{i} = S_{i} + \mu^{\alpha}_{tot,i} = S_{i} + \mu^{\alpha} +\Delta\mu^{\alpha} + \delta\mu^{\alpha}_{i} 
\end{equation}
\noindent
and the overall goodness of fit compared to the observed light curve $m^{\alpha}_{i}$ with uncertainties $\sigma_{\alpha,i}$ is 
\begin{equation}
\chi^{2} = \sum_{\alpha}\sum_{i}\left(\frac{m_{i}^{\alpha}-S_{i}-\mu_{tot,i}^{\alpha}}{\sigma_{\alpha,i}}\right)^{2}.
\end{equation}
\noindent
Given a set of trial microlensing light curves $\delta\mu_{i}^{\alpha}$, one can calculate an estimate of the source light curve $S_{i}$ by solving $\partial \chi^{2}/\partial S_{i} = 0$. In the K04 algorithm, this estimate of the source variability is then used to simplify the $\chi^{2}$ for the microlensing analyses. Here, we additionally used the DRW model to evaluate the likelihood $P(S_{i}|\hat{\sigma},\tau)$ of this source light curve model.

We ran $N=10^{6}$ trial light curves for each lens model (ten total) and for each magnification map (eight total) and set a threshold of $\chi^{2}/N_{dof} < 2$ for saving the trial outputs. Light curves with larger $\chi^{2}$ values contribute negligibly towards the final Bayesian integrals as their contributions decrease exponentially as exp($-\chi^{2}/2$) and the number of degrees of freedom $N_{dof}$ = 2105 is large. In order to obtain a non-trivial number of light curves that satisfy this threshold, we imposed an error floor of 0.05 mag for the ABC light curves and a floor of 0.07 mag for the D light curve. The larger error floor for the D light curve is due to its overall worse photometry. This results in saving approximately 0.2\% of each of the $N=10^{6}$ random trials, resulting in $\sim$ 160,000 final outputs. Further technical details of the light curve generation and fitting procedures can be found in K04 and \cite{Poindexter2010}. For each trial light curve that passes the $\chi^{2}$ threshold, we calculated the DRW likelihoods of the source light curve model for values of $\hat{\sigma}$ ranging from $\hat{\sigma}_{int}-0.15$ to $\hat{\sigma}_{int}+0.05$ in uniform increments of 0.025, while fixing $\tau$ to 3000 days. 

The parameter estimation process follows K04, where we used Bayes theorem to estimate any quantity of interest. We assumed logarithmic priors for the DRW $\hat{\sigma}$ parameter and the scaled source velocity ($P(\hat{\sigma}) \propto 1/\hat{\sigma}$ and $P(\hat{v}_{e}) \propto 1/\hat{v}_{e}$). For the scaled source size, we assumed either a linear prior, $P(\hat{R}_{s}) \propto$ constant prior, or a logarithmic prior, $P(\hat{R}_{s}) \propto 1/\hat{R}_{s}$. We also compared results with and without a uniform prior on the mean stellar mass over the range 0.1 $h^{2}M_{\odot} < \langle M_* \rangle < 1 \mbox{ }h^{2}M_{\odot}$.  

\section{Results}
\label{sec:results}
\begin{figure}
	\centering
    \includegraphics[width=0.45\textwidth]{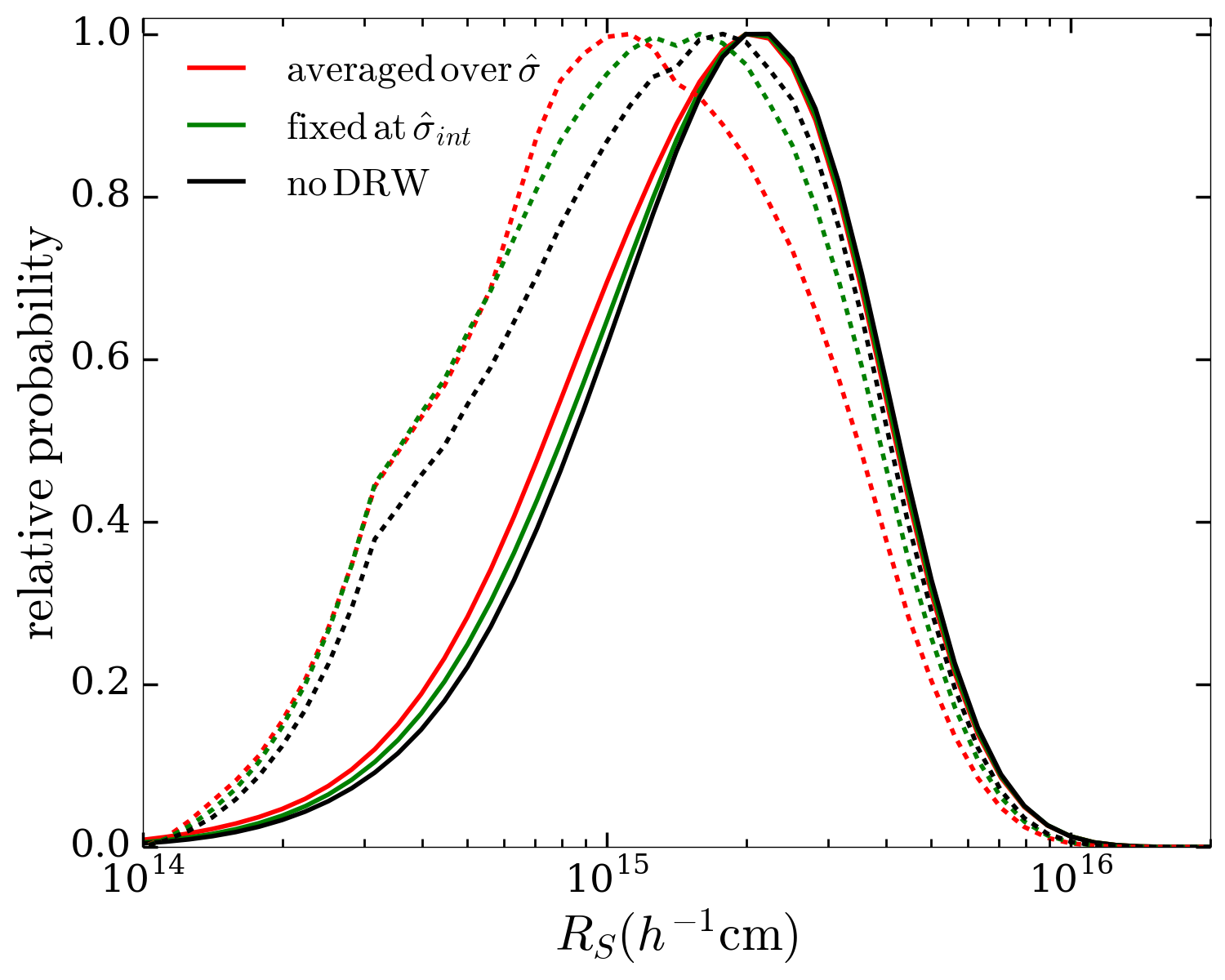}
    \caption{Relative probability distributions for the physical source size. The standard case with a log size prior and no mass prior is shown by the solid lines, while the dashed lines include a mass prior of $0.1 h^{2}M_{\odot} < \langle M_* \rangle < 1 h^{2}M_{\odot}$. The black curves include no constraints from the DRW model, while the red (green) curve is marginalized over $\hat{\sigma}$ (uses a fixed $\hat{\sigma}$).}
    \label{fig:rs}
     \includegraphics[width=0.45\textwidth]{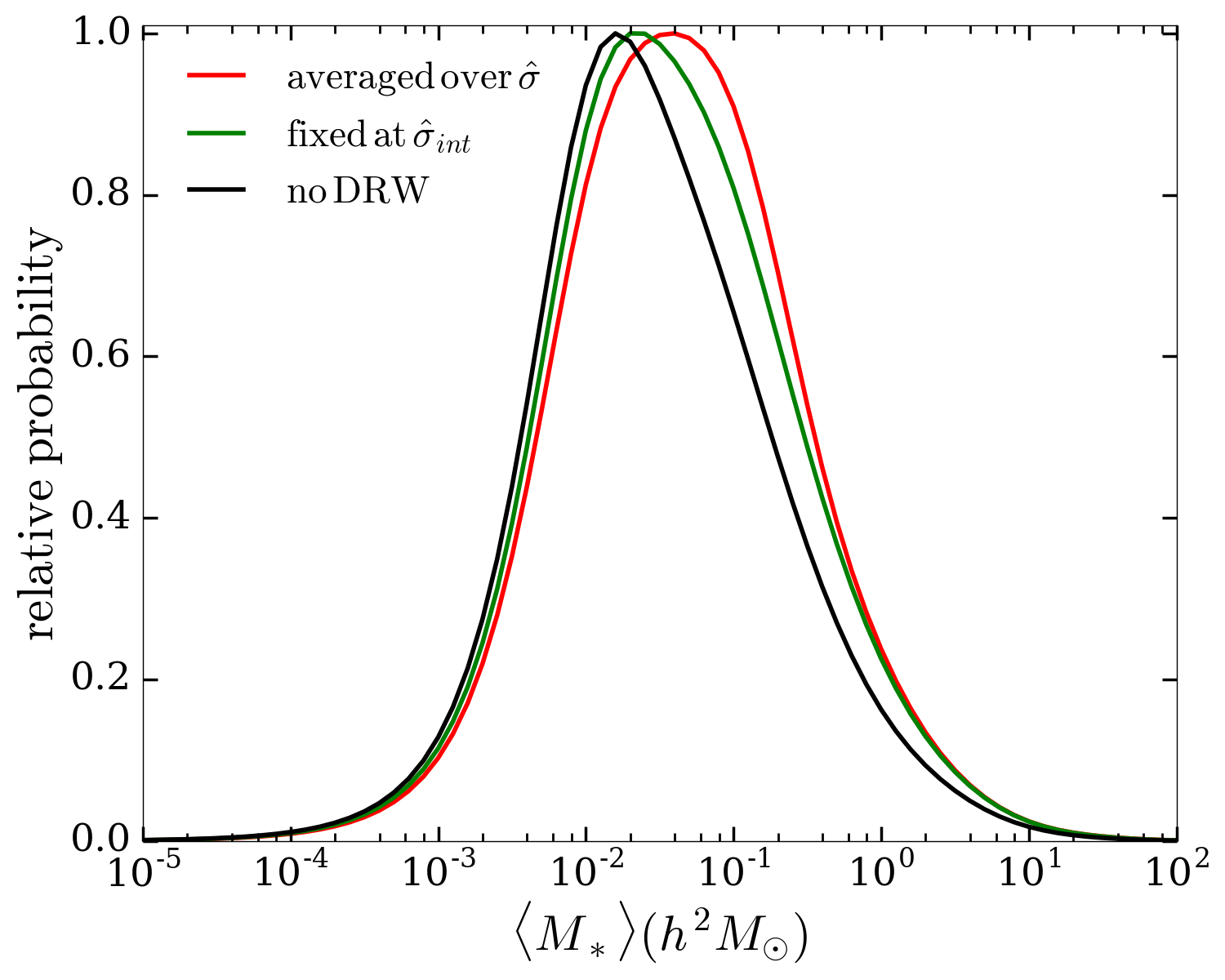}
    \caption{Relative probability distributions for the average lens mass $\langle M_* \rangle$, with the same color codings as Figure \ref{fig:rs}.}
    \label{fig:mass}
\end{figure}

\begin{figure}
	\centering
    \includegraphics[width=0.45\textwidth]{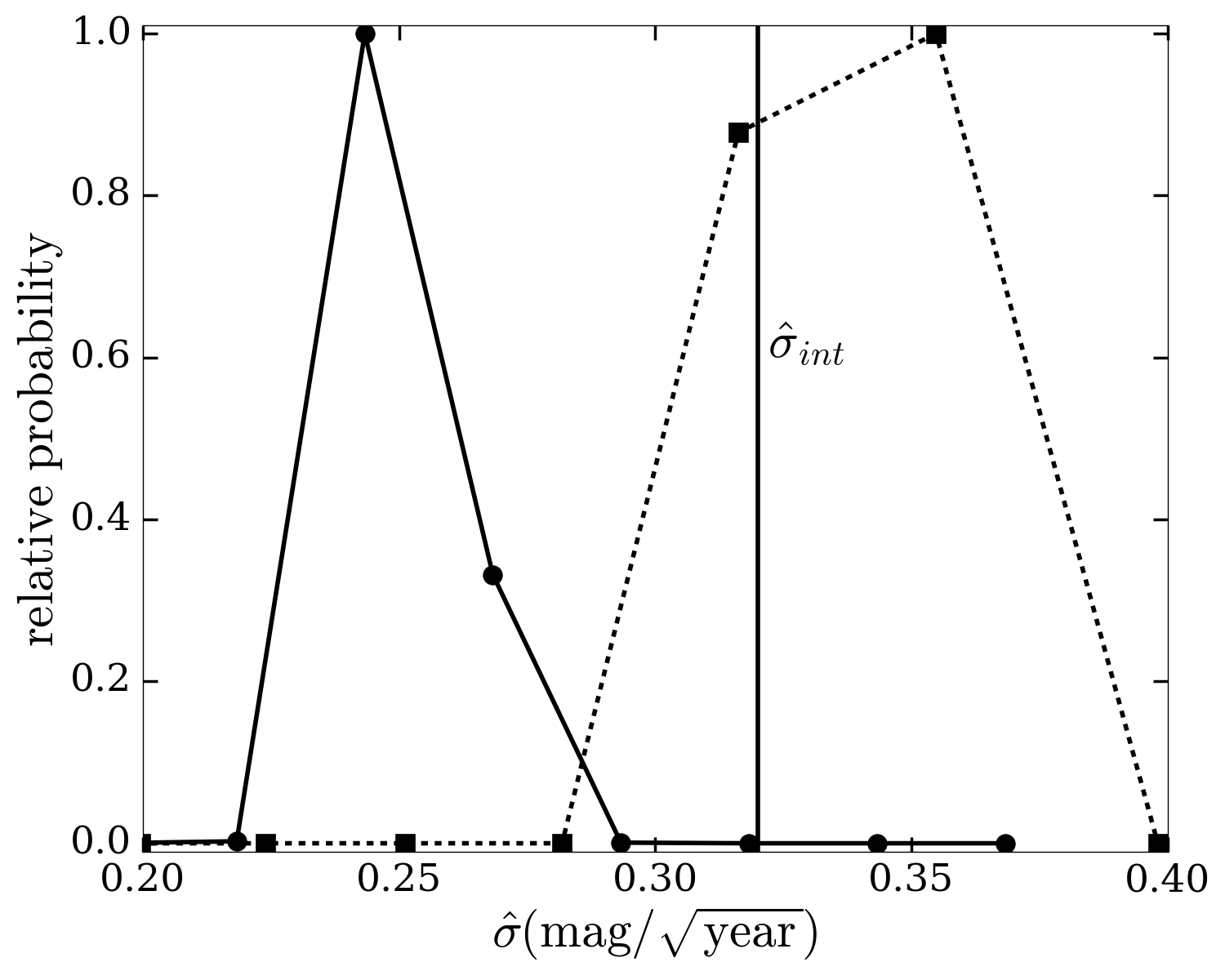}
    \caption{Relative probability distribution for the DRW $\hat{\sigma}$ parameter. The dashed line shows the results from fitting the light curves without accounting for microlensing (Figure~\ref{drwcont} with $\tau = 3000$ days) while the solid line shows the posterior estimate of $\hat{\sigma}$ from the microlensing models. As expected, some of the variability is now due to microlensing so $\hat{\sigma}$ shifts to a smaller value. The vertical line marks the value of $\hat{\sigma}_{int}$.}
    \label{fig:sigma}
\end{figure}

We calculated the relative probabilities for the source size $R_S$ and the average stellar mass $\langle M_* \rangle$ for three cases: (i) without the DRW likelihoods on the source model, (ii) with the DRW likelihoods marginalized over the range of $\hat{\sigma}$ values, and (iii) with the DRW likelihoods for a fixed $\hat{\sigma}$ value at $\hat{\sigma}_{int}$. Treating the logarithmic size prior and no mass prior as the standard case, Figures \ref{fig:rs} and \ref{fig:mass} show the probability distributions for the physical source size $R_S$ and the average lens masses $\langle M_* \rangle$. Figure \ref{fig:sigma} shows the probability distribution for $\hat{\sigma}$, marginalized over all other variables. We also included the results when we imposed a prior on the stellar mass within the range $0.1 h^{2}M_{\odot} < \langle M_* \rangle < 1 h^{2}M_{\odot}$. Table \ref{tab:results} presents a summary of our results for the standard case, where the values in brackets refer to cases that include a mass prior. 

We obtained a median log($R_{S}$/cm) = 15.26 (14.92$-$15.54 at 68\% confidence) without the DRW likelihoods, compared to a median log($R_{S}$/cm) = 15.23 (14.87$-$15.52 at 68\% confidence) with the DRW model averaged over $\hat{\sigma}$ and log($R_{S}$/cm) = 15.25 (14.89$-$15.53 at 68\% confidence) with the DRW likelihoods for a fixed $\hat{\sigma} = \hat{\sigma}_{int}$. These size estimates are consistent with each other and with the size estimate from \cite{Dai2010} of log($R_{S}$/cm) = 15.11 (14.89$-$15.32 at 68\% confidence). The median size estimates obtained from using a linear size prior are $\sim$ 0.2 dex larger than those obtained using a log size prior, but adding the DRW model likelihoods again has no significant impact on the estimated source size. Figure \ref{fig:rs} shows the source size probability distribution. 

We obtained a median estimate of the average stellar mass of 0.026 $h^{2}M_{\odot}$ (0.0052$-$0.19 at 68\% confidence) without the DRW likelihoods, compared to 0.041 $h^{2}M_{\odot}$ (0.0072$-$0.28 at 68\% confidence) when we averaged over $\hat{\sigma}$ and 0.035 $h^{2}M_{\odot}$ (0.0063$-$0.26 at 68\% confidence) using $\hat{\sigma}_{int}$. Figure \ref{fig:mass} shows the probability distribution for the average lens mass. For a standard IMF and an old stellar population, we expect $\langle M_* \rangle$ $\sim$ 0.3$M_{\odot}$ including remnants, which is broadly consistent with this result. 

The posterior probability distribution for $\hat{\sigma}$ of the source variability model is shown in Figure~\ref{fig:sigma}. The median estimate of $\hat{\sigma}$ is 0.25 mag/$\sqrt[]{\mbox{year}}$ (0.23$-$0.27 at 68\% confidence), which is approximately 1.3 times lower than the estimate of $\hat{\sigma}_{int}$ = 0.32 found for the observed light curves for the same fixed $\tau = 3000$ days. It is not surprising that the estimate of $\hat{\sigma}$ including microlensing is smaller than the estimates made from the light curves without correcting for microlensing. Since some of the observed variability is now due to microlensing rather than being intrinsic to the source, we should see a shift of $\hat{\sigma}$ to lower values. Because it is difficult to estimate $\tau$, we simply held it fixed for this experiment. Unlike the expected shift in $\hat{\sigma}$, there is also no obvious expected change in $\tau$ between the models with and without microlensing. 

\begin{table*} 
\centering
\small
\begin{threeparttable}
\caption{Microlensing parameters for RX J1131-1231} 
\centering 
\begin{tabular}{c c c c}
\hline\hline 
Parameter & No DRW & DRW $\hat{\sigma}$ marginalized & DRW fixed $\hat{\sigma}$ \\
\hline  
\rule{0pt}{2.7ex}
log($R_{S}$/cm) & 15.26$^{+0.28}_{-0.34}$ (15.13$^{+0.35}_{-0.43}$) & 15.23$^{+0.29}_{-0.36}$ (15.05$^{+0.37}_{-0.40}$) & 15.25$^{+0.28}_{-0.36}$ (15.09$^{+0.36}_{-0.42}$)\\[0.8ex]
log($\langle M_* \rangle$/$h^{2} M_{\odot}$) & $-$1.59$^{+0.87}_{-0.70}$  & $-$1.38$^{+0.83}_{-0.77}$ & $-$1.45$^{+0.87}_{-0.75}$ \\[0.5ex]

\hline
\end{tabular}
\begin{tablenotes}
      \small
      \item The median values and the errors based on the 68\% confidence interval, for cases with a log size prior and without a mass prior (the standard case). Values in brackets are when a mass prior is enforced. 
    \end{tablenotes}
\end{threeparttable}
\label{tab:results}
\end{table*}

\section{Conclusion}
\label{sec:conclusion}
The original Bayesian Monte Marlo method of K04 for analyzing quasar microlensing light curves does not attempt to evaluate the model for the intrinsic variability of the quasar. Rather it absorbs the quasar variability into the microlensing models by describing it as the average difference between the microlensing variability and the trial light curves. This process could imply atypical quasar variability and allowing such cases might affect the estimated microlensing parameters. 

To test this, we modified the initial K04 algorithm by including the DRW model to evaluate the likelihood of the source variability model. We applied the modified algorithm to the lensed quasar RX J1131$-$1231 using the $R$ band optical light curves from \cite{Tewes2013} that span almost nine years. We estimated the physical size of the quasar disk $R_S$ and the average stellar mass $\langle M_* \rangle$ for three cases: (i) without using the DRW model, (ii) using the DRW model and averaging over a range of $\hat{\sigma}$ values, and (iii) using the DRW model with a fixed $\hat{\sigma} = \hat{\sigma}_{int}$ value estimated from the observed light curves. 

We found that adding the DRW model has negligible effects on the physical parameters and that the results are consistent with those obtained without a DRW model. The likely explanation for the minimal effect is that an unlikely source model is produced only when the microlensing fits are already poor and so already contribute little statistical weight to the posterior probability distributions. At least for RX J1131$-$1231, adding the information on the likelihood of the source light curve does not significantly impact the estimate of the source size. Therefore, there appears to be no reason to undertake the challenging task of parallelizing the DRW likelihood analysis to work with the more complex ``moving stars'' version of the K04 algorithm developed by \cite{Poindexter2010}. More generally, it seems unlikely that priors on the source variability are needed for microlensing analyses. 

\section*{Acknowledgments}
\noindent
This research was supported by HST program GO-13113 provided by NASA through a grant from the Space Telescope Science Institute, which is
operated by the Association of Universities for Research in Astronomy, Inc., under NASA contract NAS5-26555. CSK is supported by NSF grant AST -1515876.  This research made use of Astropy, a
community-developed core Python package for Astronomy
(Astropy Collaboration, 2013).


\begin{thebibliography}{}
\bibitem[Andrae et al.(2013)]{Andrae2013} Andrae, R., Kim, D.-W., Bailer-Jones, \& C.~A.~L. \ 2013, \aap, 554, A137
\bibitem[Blackburne et al.(2006)]{r1131_blackburne2006} Blackburne, J.~A., Pooley, D., \& Rappaport, S. \ 2006, \apj, 640, 569
\bibitem[Blackburne et al.(2011)]{Blackburne2011} Blackburne, J.~A., Pooley, D., Rappaport, S., Schechter, P.~L. \ 2011, \apj, 729, 34
\bibitem[Chartas et al.(2009)]{r1131_chartas2009} Chartas, G., Kochanek, C.~S., Dai, X., Poindexter, S., \& Garmire, G. \ 2009, \apj, 693, 174
\bibitem[Chartas et al.(2012)]{Chartas2012} Chartas, G., Kochanek, C.~S., Dai, X., et al. \ 2012, \apj, 757, 137
\bibitem[Chartas et al.(2016a)]{r1131_chartas2016} Chartas, G., Rhea, C., Kochanek, C., et al. \ 2016, Astronomische Nachrichten, arXiv:1509.05375
\bibitem[Chartas et al.(2016b)]{r1131_chartas2016_2} Chartas, G., Krawczynski, H., Zalesky, L., et al. \ 2009, \apj, 693, 174
\bibitem[Claeskens et al.(2006)]{r1131_claeskens2006} Claeskens, J.-F., Sluse, D., Riaud, P., \& Surdej, J. \ 2006, \aap, 451, 865
\bibitem[Collier et al.(1998)]{Collier1998} Collier, S.~J., Horne, K., Kaspi, S., et al. \ 1998, \apj, 500, 162
\bibitem[Dai et al.(2010)]{Dai2010} Dai, X., Kochanek, C.~S., Chartas, G., et al. \ 2010, \apj, 709, 278
\bibitem[Fabian et al.(2002)]{Fabian2002} Fabian, A.~C., Vaughan, S., Nandra, K., et al. \ 2002, \mnras, 335, L1
\bibitem[Fausnaugh et al.(2016)]{Fausnaugh2016} Fausnaugh, M.~M., Denney, K.~D., Barth, A.~J., et al.\ 2016, \apj, 821, 56
\bibitem[Iwasawa et al.(1996)]{Iwasawa1996} Iwasawa, K., Fabian, A.~C., Reynolds, C.~S., et al. \ 1996, \mnras, 282, 1038
\bibitem[Iwasawa et al.(2004)]{Iwasawa2004} Iwasawa, K., Lee, J.~C., Young, A.~J., Reynolds, C.~S., \& Fabian, A.~C. \ 2004, \mnras, 347, 411
\bibitem[Kasliwal et al.(2015)]{Kasliwal2015} Kasliwal, V.~P., Vogeley, M.~S., \& Richards, G.~T. \ 2015, \mnras, 451, 4328
\bibitem[Kelly et al.(2009)]{Kelly2009} Kelly, B.~C., Bechtold, J., \& Siemiginowska, A. \ 2009, \apj, 698, 895
\bibitem[Kelly et al.(2011)]{Kelly2011} Kelly, B.~C., Sobolewska, M., \& Siemiginowska, A. \ 2011, \apj, 730, 52
\bibitem[Kochanek(2004)]{Kochanek2004} Kochanek, C.~S. \ 2004, \apj, 605, 58
\bibitem[Kozlowski et al. (2010)]{Kozlowski2010} Koz{\l}owski, S., Kochanek, C.~S., Udalski, A., {Wyrzykowski}, {\L}. \& the {OGLE Collaboration} \ 2010, \apj, 708, 927
\bibitem[Kozlowski (2016)]{Kozlowski2016} Koz{\l}owski, S. \ 2016, \mnras, 459, 2787
\bibitem[Kozlowski (2017)]{Kozlowski2017} Koz{\l}owski, S. \ 2017, \aap, 597, A128
\bibitem[Krawczynski \& Chartas(2016)]{r1131_kraw2016} Krawczynski, H. \& Chartas, G. \ 2016, ArXiv e-prints, arXiv:1610.06190
\bibitem[MacLeod et al. (2010)]{MacLeod2010} MacLeod, C.~L., Ivezi{\'c}, {\v Z}., {Kochanek}, C.~S., et al. \ 2010, \apj, 721, 1014
\bibitem[MacLeod et al.(2015)]{MacLeod2015} MacLeod, C.~L., Morgan, C.~W., Mosquera, A., et al. \ 2015, \apj, 806, 258
\bibitem[Morgan et al.(2006)]{r1131_morgan2006} Morgan, N.~D., Kochanek, C.~S., Falco, E.~E., \& Dai, X. \ 2006, ArXiv Astrophysics e-prints, astro-ph/0605321
\bibitem[Morgan et al.(2008)]{Morgan2008} Morgan, C.~W., Kochanek, C.~S., Dai, X., Morgan, N.~D., \& Falco, E.~E. \ 2008, \apj, 689, 755
\bibitem[Mosquera \& Kochanek (2011)] {Mosquera2011} Mosquera, A.~M. \& {Kochanek}, C.~S. \ 2011, \apj, 738, 96
\bibitem[Mosquera et al.(2013)] {Mosquera2013} Mosquera, A.~M., Kochanek, C.~S., Chen, B., et al. \ 2013, \apj, 769, 53
\bibitem[Mushotzky et al.(2011)]{Mushotzky2011} Mushotzky, R.~F., Edelson, R., Baumgartner, W., \& Gandhi, P. \ 2011, \apjl, 743, L12
\bibitem[Neronov et al.(2016)]{r1131_neronov2016} Neronov, A., Malyshev, D., \& Walter, R. \ 2016, arXiv, 1602.07601
\bibitem[Paczynski(1996)]{Paczynski1996} Paczynski, B. \ 1996, \araa, 34, 419
\bibitem[Poindexter et al.(2008)] {Poindexter2008} Poindexter, S., Morgan, N., \& Kochanek, C.~S. \ 2008, \apj, 673, 34
\bibitem[Poindexter \& Kochanek(2010)] {Poindexter2010} Poindexter, S., Morgan, N., \& Kochanek, C.~S. \ 2010, \apj, 712, 658
\bibitem[Pooley et al.(2007)]{Pooley2007} Pooley, D., Blackburne, J.~A., Rappaport, S., \& Schechter, P.~L. \ 2007, \apj, 661, 19
\bibitem[Scargle(1981)]{Scargle1981} {Scargle}, J.~D. \ 1981, \apjs, 45, 1
\bibitem[Sergeev et al.(2005)]{Sergeev2005} Sergeev, S.~G., Doroshenko, V.~T., Golubinskiy, Y.~V., Merkulova, N.~I., \& Sergeeva, E.~A. \ 2005, \apj, 622, 129
\bibitem[Shakura \& Sunyaev(1973)]{Shakura1973} Shakura, N.~I., \& Sunyaev, R.~A.\ 1973, \aap, 24, 337 
\bibitem[Shappee et al.(2014)]{Shappee2014} Shappee, B.~J., \& Prieto, J.~L., \& Grupe, D., et al.\ 2014, \apj, 788, 48
\bibitem[Sluse et al.(2003))]{Sluse2003} Sluse, D., Surdej, J., Claeskens, J.-F., et al. \ 2003, \aap, 406, L43
\bibitem[Tanaka et al.(1995)]{Tanaka1995} Tanaka, Y., Nandra, K., Fabian, A.~C., et al. \ 1995, \nat, 375, 659
\bibitem[Tewes et al.(2013)]{Tewes2013} Tewes, M., Courbin, F., Meylan, G., et al.,\ 2013, \aap, 556, A22
\bibitem[Wambsganss(2006)]{Wambsganss2006} Wambsganss, J.\ 2006, Saas-Fee Advanced Course 33: Gravitational Lensing: Strong, Weak and Micro, 453 
\bibitem[Zu et al.(2013)]{Zu2013} Zu, Y., Kochanek, C.~S., Koz{\l}owski, S., \& Udalski, A. \ 2013, \apj, 765, 106

\end{thebibliography}
\end{document}